\documentclass[times,twocolumn,final]{elsarticle}

\usepackage{medima}
\usepackage{framed,multirow}

\usepackage{amssymb}
\usepackage{amsmath}
\usepackage{latexsym}

\usepackage{url}
\usepackage{xcolor}
\usepackage{booktabs}
\usepackage{hyperref}
\hypersetup{
  pdfinfo={
  pdfproducer={},
  Title={},
  Subject={},
  Author={},
  }
}
\usepackage{siunitx}
\usepackage{graphicx}

\definecolor{newcolor}{rgb}{.8,.349,.1}

\journal{Medical Image Analysis}

\begin{document}

\verso{Xueqi Guo \textit{et~al.}}

\begin{frontmatter}

\title{TAI-GAN: A Temporally and Anatomically Informed Generative Adversarial Network for early-to-late frame conversion in dynamic cardiac PET inter-frame motion correction}%

\author[1]{Xueqi \snm{Guo}\corref{cor1}}
\ead{xueqi.guo@yale.edu}
\author[2]{Luyao \snm{Shi}} 
\author[1]{Xiongchao \snm{Chen}}
\author[1]{Qiong \snm{Liu}}
\author[1]{Bo \snm{Zhou}}
\author[1]{Huidong \snm{Xie}}
\author[3]{Yi-Hwa \snm{Liu}}
\author[4]{Richard \snm{Palyo}}
\author[1,3,5]{Edward J. \snm{Miller}}
\author[1,3,5]{Albert J. \snm{Sinusas}}
\author[1,5]{Lawrence \snm{Staib}}
\author[6]{Bruce \snm{Spottiswoode}}
\author[1,5]{Chi \snm{Liu}\corref{cor1}}
\ead{chi.liu@yale.edu}
\author[1,5]{Nicha C. \snm{Dvornek}\corref{cor1}}
\ead{nicha.dvornek@yale.edu}
\cortext[cor1]{Corresponding authors.}
  
\address[1]{Department of Biomedical Engineering, Yale University, New Haven, CT, USA.}
\address[2]{IBM Research, San Jose, CA, USA.}
\address[3]{Department of Internal Medicine, Yale University, New Haven, CT, USA}
\address[4]{Yale New Haven Hospital, New Haven, CT, USA.}
\address[5]{Department of Radiology and Biomedical Imaging, Yale School of Medicine, New
Haven, CT, USA}
\address[6]{Siemens Medical Solutions USA, Inc., Knoxville, TN, USA.}

\received{xx xxx 2023}
\finalform{xx xxx 2023}
\accepted{xx xxx 2023}
\availableonline{xx xxx 2023}
\communicated{xxx}

\begin{abstract}
Inter-frame motion in dynamic cardiac positron emission tomography (PET) using rubidium-82 ($^{82}$Rb) myocardial perfusion imaging impacts myocardial blood flow (MBF) quantification and the diagnosis accuracy of coronary artery diseases. However, the high cross-frame distribution variation due to rapid tracer kinetics poses a considerable challenge for inter-frame motion correction, especially for early frames where intensity-based image registration techniques often fail. To address this issue, we propose a novel method called Temporally and Anatomically Informed Generative Adversarial Network (TAI-GAN) that utilizes an all-to-one mapping to convert early frames into those with tracer distribution similar to the last reference frame. The TAI-GAN consists of a feature-wise linear modulation layer that encodes channel-wise parameters generated from temporal information and rough cardiac segmentation masks with local shifts that serve as anatomical information. Our proposed method was evaluated on a clinical $^{82}$Rb PET dataset, and the results show that our TAI-GAN can produce converted early frames with high image quality, comparable to the real reference frames. After TAI-GAN conversion, the motion estimation accuracy and subsequent myocardial blood flow (MBF) quantification with both conventional and deep learning-based motion correction methods were improved compared to using the original frames.

\end{abstract}


\end{frontmatter}



\section{Introduction}
Dynamic cardiac positron emission tomography (PET) myocardial perfusion imaging is more accurate in detecting coronary artery disease compared to other non-invasive imaging procedures \citep{prior2012quantification}. A dynamic frame sequence is obtained after the injection of the radioactive tracer rubidium-82 ($^{82}$Rb), which lasts several minutes until the myocardium is sufficiently perfused. Regions of interest (ROIs) of the myocardium tissue and left ventricle blood pool (LVBP) are labeled from the reconstructed frames to collect time-activity curves (TACs) for the following kinetic modeling and the estimation of myocardial blood flow (MBF) and myocardial flow reserve (MFR). MBF and MFR quantification have demonstrated improved diagnostic and prognostic effectiveness \citep{sciagra2021eanm}.

However, in dynamic cardiac PET, subject motion originating from respiratory, cardiac, and voluntary body movements can seriously impact both voxel-based and ROI-based MBF quantification \citep{hunter2016patient}. On the one hand, the intra-frame motion can cause blurriness in reconstructed frames and incorrect activity measurements. On the other hand, the inter-frame motion introduces spatial mismatch across the dynamic frames, resulting in attenuation mismatch, incorrect TAC measurements, and subsequent MBF estimation errors. Current inter-frame motion correction approaches for dynamic PET include external motion tracking systems \citep{noonan2015repurposing,lu2018respiratory}, data-driven motion estimation algorithms \citep{ren2017data,feng2017self,lu2019data,lu2020data}, and non-rigid registration using conventional optimization \citep{mourik2009off,jiao2014spatio,guo2022inter,sun2022iterative} or deep learning-based methods \cite{zhou2021mdpet,guo2022unsupervised,guo2022mcp,zhou2023fast,guo2023mcp}. However, few studies have specifically addressed the challenges in inter-frame motion correction for cardiac dynamic PET with the rapid tracer kinetics of $^{82}$Rb. In the early phases of tracer perfusion, $^{82}$Rb concentrates in the right ventricle blood pool (RVBP) first and subsequently in LVBP; in the late scans, $^{82}$Rb is well distributed in myocardial tissue. This significant change in tracer distribution can substantially complicate inter-frame motion correction since intensity-based frame registration typically depends on the resemblance between the two frames that are to be registered, especially between the early and late frames \citep{lu2020patient,shi2021automatic}. In $^{82}$Rb dynamic cardiac PET, most existing motion correction research and clinical software focuses exclusively on the later frames in the myocardial perfusion stage \citep{burckhardt2009cardiac,woo2011automatic,rubeaux2017enhancing,lu2020patient}. A blood pool isolation strategy was proposed for the blood pool phase \citep{lee2016image} but not for the late myocardium phase frames. A frame registration method using normalized gradient fields instead of original intensities was proposed to narrow the gap, but unidentifiable boundaries in the transition phase frames and the usage of the blurred summed tissue phase frame as the reference might introduce additional errors to motion estimation \citep{lee2020automated}. A motion correction framework under supervised learning was proposed for $^{82}$Rb dynamic cardiac PET under simulated translational motion \citep{shi2021automatic}, but the network requires the training of two separate models to address the variation between early and late frames, which can be computationally expensive and inconvenient in clinical practice. 

An alternative solution is using frame conversion to generate mapped early frames that appear similar in tracer distribution to the corresponding late frame to assist standard motion correction methods. Recent studies have indicated that utilizing modality conversion through image synthesis can enhance optimization in multi-modality image registration by simplifying it to intra-modality \citep{iglesias2013synthesizing,xiao2020review}. For instance, image synthesis methods have been utilized to convert magnetic resonance (MR) to computed tomography (CT) images \citep{roy2014mr,cao2018region}, MR T1-weighted (T1w) to T2-weighted (T2w) images \citep{chen2017cross,liu2019image}, and MR to X-ray mammography \citep{maul2021x}; however, few studies involve PET images.

Such medical image synthesis tasks have largely utilized convolutional neural networks (CNNs) \citep{sevetlidis2016whole,liu2019image,zhou2020hi}. In molecular imaging, CNNs have been successfully implemented in the generation of attenuation maps \citep{shi2019novel,shi2022deep,chen2022direct}, cross-tracer images \citep{wang2021generation}, and parametric $K_i$ images \citep{miao2023generation}, mostly deploying the structure of a 3-D U-Net \citep{cciccek20163d}. Generative adversarial networks (GANs) \citep{goodfellow2020generative} and their derivatives are a group of CNNs including a generator and a discriminator which are trained in a process of competition under an adversarial loss. GAN-based image synthesis from MR T1w images has been implemented to generate CT images \citep{lei2019mri,abu2021paired}, T2w images \citep{dar2019image}, and fluid-attenuated inversion recovery images \citep{yu20183d}. In nuclear medicine imaging, recent works have investigated attenuation map generation for single-photon emission computed tomography (SPECT) \citep{shi2020deep} and the direct SPECT attenuation correction deploying conditional GANs (cGANs), a GAN variation with conditional restraint for specific mapping. A 3-D cGAN with self-attention and spectral normalization mechanism was proposed to synthesize brain PET from MR in an Alzheimer's disease database \citep{lan2021three}. A cGAN with visual information fidelity loss was developed for generating synthetic CT images for the attenuation correction of small animal PET \citep{li2021small}. A short-to-long acquisition conversion module using a cGAN was included in a motion correction and reconstruction framework for accelerated PET \citep{zhou2023fast}.

A closely related work to improve inter-frame motion correction in dynamic PET deployed a cGAN to convert low-count early frames to the high-count late frame of brain \citep{sundar2021conditional} and total-body dynamic scans \citep{sundar2021data}. In this way, the low-count limitation and the different tracer uptake patterns of early frames are addressed, and the motion correction using the standard multi-scale mutual information method is successfully improved. However, this approach involves training one-to-one mappings between each specific early frame and the reference frame, which may not generalize well to new acquisitions and can be challenging to implement in clinical practice settings. Additionally, the backbone of the generator is a simple 3-D U-Net, and the tracer kinetics and related temporal analysis are not incorporated in network training. This method was originally developed on 2-deoxy-2-[$^{18}$F]fluoro-D-glucose (FDG) PET scans, while $^{82}$Rb kinetics is substantially more rapidly changing than FDG. This is a potential limitation of their approach when directly applied to $^{82}$Rb scans. Recently, feature-wise linear modulation (FiLM) \citep{perez2018film} has been reported to be effective in encoding conditional information and vision reasoning. FiLM layers have been incorporated into GAN models to encode text or semantic information for natural images, specifically fashion image generation \citep{ak2019semantically,mao2019bilinear,ak2020semantically}. In MRI generation, the FiLM layer was proposed to process metadata \citep{rachmadi2019predicting}. Similarly, \cite{dey2021generative} proposed to use the FiLM layer as a conditional embedding in a GAN for deformable template generation in MRI registration. However, such a structure has not been proposed either for tracer kinetics information encoding or in dynamic PET synthesis.

\begin{figure*}[!t]
\centering
\includegraphics[width=\textwidth]{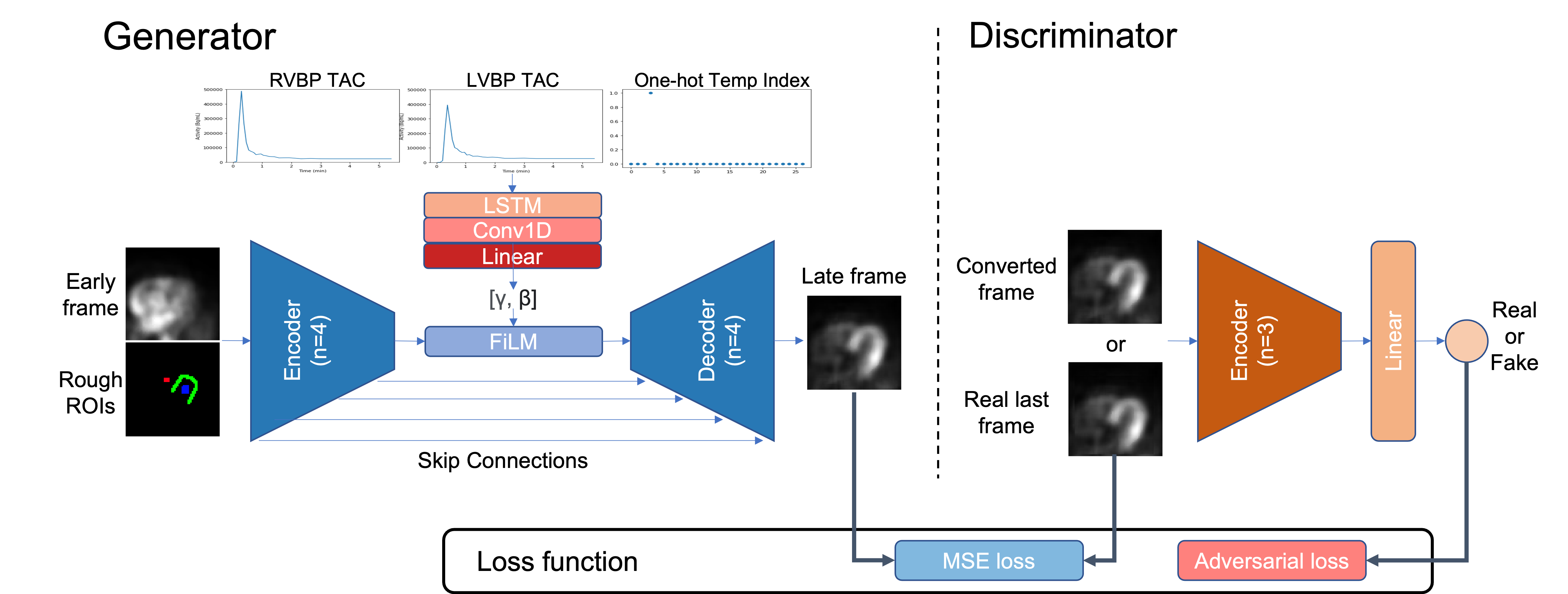}
\caption{The architecture of the proposed temporally and anatomically informed generative adversarial network (TAI-GAN).}
\label{network}
\end{figure*}

Thus, in this work, we introduce a new framework named Temporally and Anatomically Informed GAN (TAI-GAN), an all-to-one mapping approach that converts all early frames into those with the appearance of the last reference frame. To provide the temporal information to the generator, we incorporate a FiLM layer that embeds channel-wise parameters generated from the temporal frame index and blood pool TACs. To inform the network with auxiliary anatomical locators, we utilize rough segmentation masks of RVBP, LVBP, and myocardium with local shifts as the additional image input channel. TAI-GAN is the first work for dynamic cardiac PET frame conversion that addresses the challenges of both high tracer distribution variability and spatial mismatch by incorporating both temporal and anatomical information, effectively handling the variations in tracer distribution over time and ensuring accurate spatial alignment in the generated frames. Using a 5-fold cross-validation, we evaluated TAI-GAN in terms of frame conversion similarity, motion correction accuracy, and MBF quantification errors.

\section{Methods}

\subsection{Clinical cardiac dataset and pre-processing}

A total of 104 anonymized clinical dynamic cardiac $^{82}$Rb PET scans from 73 subjects at Yale New Haven Hospital were included in this study, with the approval of the Yale Institutional Review Board. Each scan was acquired using a GE Discovery 690 PET/CT scanner after $^{82}$Rb generation using a Bracco
Diagnostics Inc.\ commercial generator and weight-based $^{82}$Rb delivery using a programmed infusion. In each scan, the list-mode data of the first 6 minutes and 10 seconds were rebinned into a sequence of 27 dynamic frames (14×5s, 6×10s, 3×20s, 3×30s, 1×90s) in total. All the frames were reconstructed into 128 × 128 × 47 volumes with a voxel size of 3.125 × 3.125 × 3.270 mm$^3$ using the ordered subset expectation maximization algorithm (OSEM, 2 iterations, and 24 subsets) with the corrections of isotope decay, attenuation, scatter, random and prompt-gamma coincidences, detector efficiency, and deadtime, and then filtered with a Butterworth filter (cutoff frequency = 21 mm$^{-1}$, order = 5). The intensities of the dynamic frames were calibrated as activity concentration (Bq/mL). With the Corridor 4DM software in the clinical setting, technologists manually performed frame-by-frame motion correction and identified each scan as either motion-free (total motion no more than 3 mm, \cite{hunter2016patient,shi2021automatic}) or with moderate to significant motion. The demographic information and characteristics of the dataset are summarized in Table \ref{table1}.


\begin{table}[htb!]
\newcommand{\tabincell}[2]{\begin{tabular}{@{}#1@{}}#2\end{tabular}}
\centering
\caption{The demographic information and characteristics of the cardiac dataset.}
\label{table1}
\resizebox{0.48\textwidth}{!}{
\begin{tabular}{@{}ccc@{}}
\toprule
Group & A (motion-free) & B (with motion)\\ \midrule
\vspace{5pt}
Scan time & Dec. 2019 to Jan. 2020 & Oct. 2022\\
\vspace{5pt}
\tabincell{c}{Number of scans\\ (rest/stress)} & 85 (55/30) & 19 (10/9)\\
\vspace{5pt}
\tabincell{c}{Number of patients\\ (female/male)} & 59 (34/25) & 14 (9/5)\\
\vspace{5pt}
\tabincell{c}{Age (year, mean $\pm$ \\ standard deviation)} & 62.9 $\pm$ 11.3 & 65.8 $\pm$ 10.9\\
\tabincell{c}{Body mass index ($kg/m^2$, \\ mean $\pm$ standard deviation)} & 37.1 $\pm$ 10.9 & 32.3 $\pm$ 7.3\\
\bottomrule
\end{tabular}
}
\end{table}

The rough segmentations of RVBP, LVBP, and LV myocardium were manually annotated on all scans using the last dynamic frame as a reference for the generation of TAC and subsequent quantification of MBF. The location of the LV inferior wall center was manually labeled for each scan on the last dynamic frame for cardiac central cropping \citep{shi2021automatic}.

\subsection{Proposed network}
The model architecture of the proposed TAI-GAN is illustrated in Figure \ref{network}. The generator maps the input early frame to its corresponding late frame, and the discriminator takes either the true or synthesized late frame as the input and encodes it to determine whether it is real or fake. The backbone structure of the generator is a 3-D U-Net with skip connections between the 4 downsample or upsample levels in both the encoder and decoder, with modifications specifically designed for auxiliary guidance in terms of temporal and anatomical information. The backbone structure of the discriminator is a PatchGAN \citep{isola2017image,sundar2021conditional} with 3 encoding levels and 1 linear output layer.

Given the significant discrepancy in the tracer distribution across the various dynamic frame phases, the model is \textit{temporally informed} by the auxiliary input of the concatenated RVBP and LVBP TACs along with the one-hot coded temporal frame index. Considering the recent success in implementing long short-term memory (LSTM) networks \citep{hochreiter1997long} in 1-D biomedical sequential data analyses \citep{guo2019deep,dvornek2017identifying,guo2022early}, we deploy an LSTM to analyze the concatenated temporal input. Subsequently, a 1-D convolutional layer and a linear layer encode the LSTM output to the manipulation parameters $\gamma$ and $\beta$ for the following channel-wise linear modulation, formulated as:
\begin{equation}
\label{film_1}
[\gamma,\beta]=f(x),
\end{equation}
where $f(\cdot)$ represents the temporal encoding module (concatenated LSTM, 1-D convolution, and linear layer), and $x$ represents the concatenated auxiliary temporal inputs (RVBP TAC, LVBP TAC, and the one-hot frame index). After this, the bottleneck feature map of the 3-D U-Net is modulated by the channel-wise FiLM operator using $\gamma$ as the scale factor and $\beta$ as the bias: 
\begin{equation}
\label{film_2}
FiLM(X_i) = \gamma_i \cdot X_i+\beta_i,
\end{equation}
where for the $i^{th}$ channel of the bottleneck feature map $X_i$, $\gamma_i$ and $\beta_i$ are respectively the scale factor and the bias of the corresponding channel, and $FiLM(\cdot)$ represents the FiLM operator.

The model is \textit{anatomically informed} by the rough segmentations of the cardiac ROIs as the anatomical locator. The rough segmentations of RVBP, LVBP, and myocardium are concatenated with the early frame as the dual-channel input of the generator. These labeled masks indicate the location of different ROIs within the dynamic frames and provide anatomical information to the generator about the location of each region. This is designed to introduce consistency and prevent spatial mismatch artifacts of the myocardium in frame conversion, which are particularly useful in distinguishing between the early RV and LV phases. Without such anatomical guidance, the conversion result of an early frame in the RV phase might have the artifact of myocardium located near the RV instead of LV due to mistaking the RVBP as LVBP. To prevent errors introduced by the misalignment between the early frame and the last frame where the ROIs are labeled, during training, random local shifts of the segmentations are deployed before concatenating with the early frame. In this way, the robustness of the model is improved under weak guidance, and the model is trained to be able to handle real-patient data when significant motion may be present.

The loss function of the TAI-GAN model contains both a voxel-wise mean squared error (MSE) loss as the image similarity penalty and the cross entropy adversarial loss in the min-max game between the generator and the discriminator. As in Eq. \ref{loss_1}-\ref{loss_3} below,
\begin{equation}
\begin{aligned} 
L_{mse} &= \frac{1}{N}\sum\limits_{n=1}^{N}(G(F_i)_n - (F_L)_n)^2,
\end{aligned} 
\label{loss_1}
\end{equation}

\begin{equation}
\begin{aligned} 
L_{adv} &= -\log(D(F_L))-\log(1-D(G(F_i))),
\end{aligned} 
\label{loss_2}
\end{equation}

\begin{equation}
\begin{aligned} 
\hat{G},\hat{D} &= arg\min_{G}\max_{D}(L_{mse}+L_{adv}),
\end{aligned} 
\label{loss_3}
\end{equation}
where $L_{mse}$ is the MSE loss, $N$ is the total number of voxels in one dynamic frame, $G$ is the generator, $G(F_i)$ is the converted late frame of the $i^{th}$ early input frame $F_i$, $F_L$ is the true last frame as the reference, $L_{adv}$ is the adversarial loss, and $D$ is the discriminator. Both the generator and the discriminator networks are trained end-to-end under a min-max game with the goal of improving the quality of frame conversion. 

\subsection{Implementation details}
All the early frames with tracer concentration greater than 10\% of the maximal activity in LVBP TAC are included in implementation to be converted to the corresponding last frames. All the earlier frames are ignored since these very early frames with LVBP activity lower than 10\% maximum have no meaningful impact on image-derived input function and consequently the MBF measurement \citep{shi2021automatic}. The EQ frame was defined as the first frame in which the LVBP activity is equal to or higher than the RVBP activity \citep{shi2021automatic} and automatically labeled based on the LVBP and RVBP TACs, and the same temporal normalization process as in \citet{shi2021automatic} was applied to all the dynamic frames. The intensity normalization of all the frames was applied individually to the range of [-1,1] prior to the conversion network input to enhance GAN optimization. The network training was based on patches with the size of (64,64,32) that are randomly cropped near the center of the LV inferior wall, randomly rotated in the xy plane in the range of [-45$^{\circ}$,45$^{\circ}$], and randomly shifted in the range of [-5,5] voxels along the x, y, and z-axis for data augmentation.

Due to the low practicality of training each one-to-one mapping for all the early frames, we trained two single-pair mappings using a vanilla GAN (3-D U-Net generator) and only the adversarial loss to compare with \citet{sundar2021conditional}. The two single-pair mappings are one frame before and after the EQ frame, which are respectively designated as EQ-1 and EQ+1. Additionally, we implemented both the vanilla GAN and the MSE loss GAN as the two baselines of all-to-one conversion. An ablation study was conducted to evaluate the effectiveness of the auxiliary temporal and anatomic information analysis.

All frame conversion models are implemented using PyTorch and trained on an NVIDIA A40 GPU with an Adam optimizer (learning rate $LR_G$=2e-4, $LR_D$=5e-5). The training and evaluation are conducted under 5-fold cross-validation on the 85 motion-free scans. Each fold contained 68 scans for training and 17 scans that were randomly left out for inference. For one-to-one mappings, the stopping epoch was 800, and for all-to-one models, the stopping epoch was set to 100.

\begin{figure*}[t]
\centering
\includegraphics[width=0.8\textwidth,keepaspectratio]{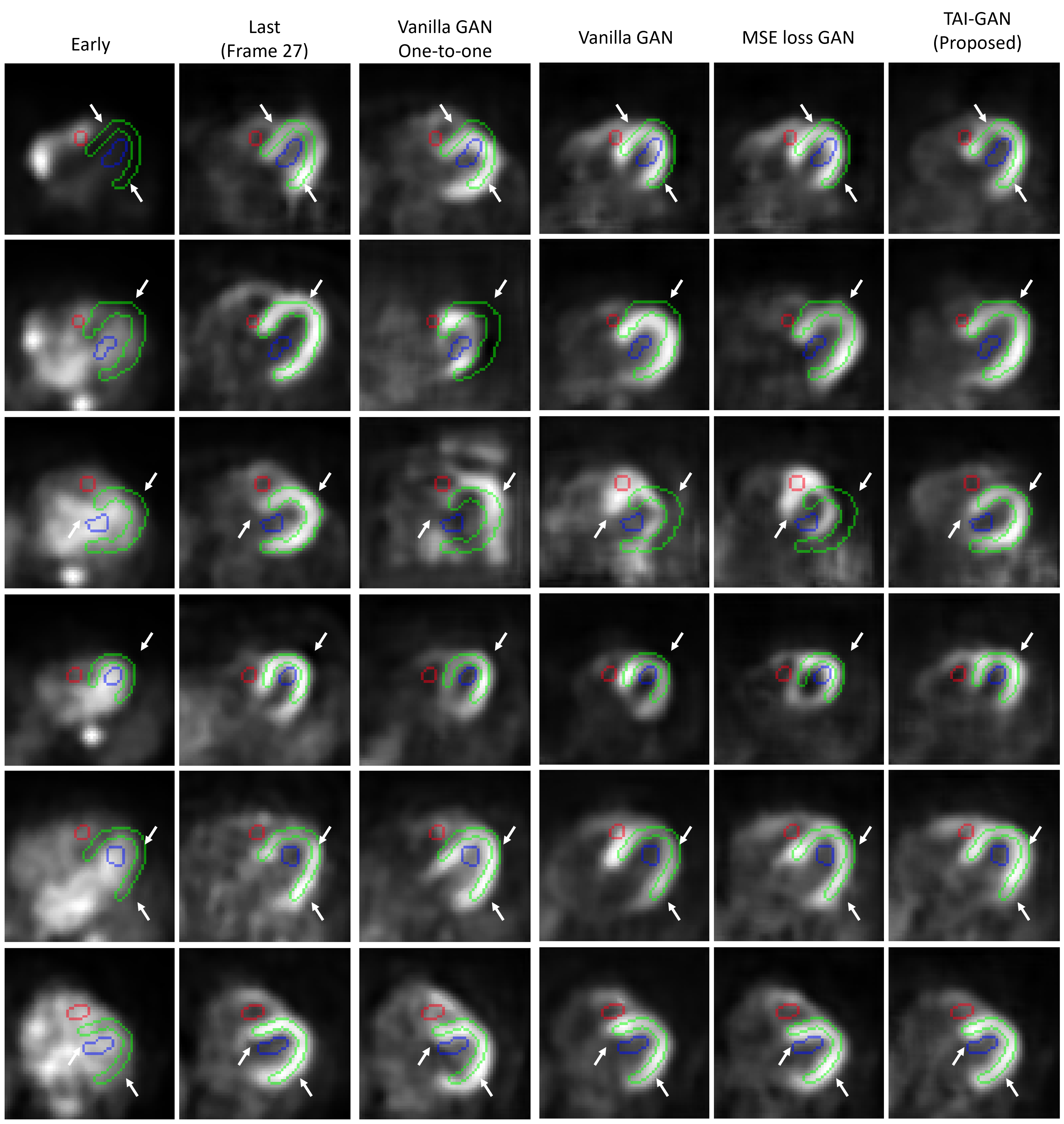}
\caption{Sample results of early-to-late frame conversion using each method with overlaid RVBP (red), LVBP (blue), and myocardium (green) segmentation contours, with arrows highlighting alignment or mismatch between the cardiac segmentations and structures.}
\label{conversion_result}
\end{figure*}

\subsection{Frame mapping evaluation}
The qualitative evaluation of frame generation results involves comparing the visualizations of the generated last frames and the real last frames given the real early frame inputs. The cardiac segmentations are overlaid on the visualized frames as locators for the anatomical regions.

The quantitative evaluations of frame generation results are the normalized mean absolute error (NMAE), MSE, structural similarity index (SSIM), and peak signal-to-noise ratio (PSNR) calculated between the real and generated last frames. As in Eq. \ref{nmae}-\ref{psnr},

\begin{equation}
\begin{aligned} 
\mathrm{NMAE} &= \frac{1}{N}\sum\limits_{n=1}^{N}\frac{|(\hat{F_L})_n - (F_L)_n|}{\max(F_L)-\min(F_L)},
\end{aligned} 
\label{nmae}
\end{equation}

\begin{equation}
\begin{aligned} 
\mathrm{MSE} &= \frac{1}{N}\sum\limits_{n=1}^{N}((\hat{F_L})_n - (F_L)_n)^2,
\end{aligned} 
\label{mse}
\end{equation}

\begin{equation}
\footnotesize
\begin{aligned} 
\mathrm{SSIM}(F_L, \hat{F_L})&=\frac{(2\mu(F_L)\mu(\hat{F_L})+C_1)(2\sigma(F_L,\hat{F_L})+C_2)}{(\mu^2(F_L)+\mu^2(\hat{F_L})+C_1)(\sigma^2(F_L)+\sigma^2(\hat{F_L})+C_2)},
\end{aligned} 
\label{ssim}
\end{equation}

\begin{equation}
\begin{aligned} 
\mathrm{PSNR} &= 10\cdot \log_{10}\left(\frac{\max(F_L)^2}{\mathrm{MSE}}\right),
\end{aligned} 
\label{psnr}
\end{equation}
where $N$ is the total number of voxels in one dynamic frame, $\hat{F_L}$ is the generated last frame, $F_L$ is the real last frame, $\max(F_L)$ and $\min(F_L)$ are respectively the maximum and minimum values of the real last frame, $max(\hat{F_L})$ and $min(\hat{F_L})$ are respectively the maximum and minimum values of the generated last frame, $\mu(F_L)$ and $\mu(\hat{F_L})$ are respectively the mean values of the real and generated last frame, $\sigma^2(F_L)$ and $\sigma^2(\hat{F_L})$ are respectively the variance of the real and generated last frame, $\sigma(F_L,\hat{F_L})$ is the covariance of $F_L$ and $\hat{F_L}$, and $C_1=(0.01\cdot DR)^2$ and $C_2=(0.03\cdot DR)^2$ are the two constants to stabilize the ratio when the denominator is weak, where $DR$ is the dynamic range of the voxel values.

\newcommand{\tabincell}[2]{\begin{tabular}{@{}#1@{}}#2\end{tabular}}
\begin{table*}[t]
\centering
\caption{Quantitative image similarity evaluation of early-to-late frame conversion (mean $\pm$ standard deviation) with the best results marked \textbf{in bold}.}
\label{frame_quan}
\resizebox{0.9\textwidth}{!}{
\begin{tabular}{c|c|c|c|c|c}
\hline
Test set & Metric & {\tabincell{c}{Vanilla GAN\\One-to-one}} & {\tabincell{c}{Vanilla \\GAN}} & {\tabincell{c}{MSE loss \\GAN}} & {\tabincell{c}{TAI-GAN\\(Proposed)}}\\
\hline
\multirow{4}{*}{EQ-1} & NMAE & 0.068 $\pm$ 0.002$^{*}$ & 0.063 $\pm$ 0.005$^{*}$ &0.059 $\pm$ 0.009 & \bfseries 0.057 $\pm$ 0.005\\

 \cline{2-6}
 & MSE & 0.057 $\pm$ 0.001$^{*}$ & 0.050 $\pm$ 0.006$^{*}$ & 0.044 $\pm$ 0.011 &\bfseries 0.040 $\pm$ 0.005\\
 \cline{2-6}
 & SSIM & 0.557 $\pm$ 0.017$^{*}$ & 0.640 $\pm$ 0.021 &0.633 $\pm$ 0.053$^{*}$ & \bfseries 0.657 $\pm$ 0.018\\
 \cline{2-6}
 & PSNR & 18.678 $\pm$ 0.116$^{*}$ & 19.370 $\pm$ 0.474$^{*}$ &19.950 $\pm$ 0.949 & \bfseries 20.335 $\pm$ 0.530\\
 \hline
 \hline
\multirow{4}{*}{EQ+1} & NMAE & 0.050 $\pm$ 0.011 & 0.053 $\pm$ 0.003 & 0.051 $\pm$ 0.006 &\bfseries 0.048 $\pm$ 0.003\\
 \cline{2-6}
 & MSE & 0.032 $\pm$ 0.014 & 0.034 $\pm$ 0.002 &0.033 $\pm$ 0.006 &\bfseries 0.032 $\pm$ 0.002\\
 \cline{2-6}
 & SSIM & 0.669 $\pm$ 0.061$^{*}$ & 0.679 $\pm$ 0.014 & 0.680 $\pm$ 0.011 & \bfseries 0.691 $\pm$ 0.013\\
 \cline{2-6}
 & PSNR & 21.323 $\pm$ 1.800 & 21.014 $\pm$ 0.355 &21.188 $\pm$ 0.757 &\bfseries 21.361 $\pm$ 0.205\\
 \hline
  \hline
\multirow{4}{*}{\tabincell{c}{All\\Pre-EQ\\frames}} & NMAE & {-} & 0.072 $\pm$ 0.007$^{*}$ & 0.066 $\pm$ 0.011 & \bfseries 0.062 $\pm$ 0.008\\
 \cline{2-6}
 & MSE & {-} & 0.063 $\pm$ 0.010$^{*}$ & 0.053 $\pm$ 0.016 & \bfseries 0.046 $\pm$ 0.009\\
 \cline{2-6}
 & SSIM & {-}  & 0.594 $\pm$ 0.012$^{*}$ &0.596 $\pm$ 0.047$^{*}$ &\bfseries 0.627 $\pm$ 0.025\\
 \cline{2-6}
 & PSNR & {-} & 18.507 $\pm$ 0.474$^{*}$ & 19.269 $\pm$ 1.036$^{*}$ & \bfseries 19.834 $\pm$ 0.738\\
 \hline
  \hline

 \multirow{4}{*}{\tabincell{c}{All \\frames}} & NMAE & {-} & 0.047 $\pm$ 0.004$^{*}$ & 0.044 $\pm$ 0.002 &\bfseries0.040 $\pm$ 0.002\\ 
 \cline{2-6}
 & MSE & {-} & 0.027 $\pm$ 0.002$^{*}$ & 0.024 $\pm$ 0.003 & \bfseries 0.021 $\pm$ 0.002\\
 \cline{2-6}
 & SSIM & {-} & 0.708 $\pm$ 0.010$^{*}$ & 0.716 $\pm$ 0.007$^{*}$ & \bfseries 0.733 $\pm$ 0.018\\
 \cline{2-6}
 & PSNR & {-} & 22.803 $\pm$ 0.530$^{*}$ & 23.241 $\pm$ 0.342$^{*}$ & \bfseries 23.799 $\pm$ 0.466\\
 \hline
\multicolumn{6}{l}{{\footnotesize  $^{*}$P$\textless 0.05$ between the current method and TAI-GAN (paired two-tailed t-test).}}\\ 
\end{tabular}
}
\end{table*}

\subsection{Simulated and real-patient motion correction}
A motion simulation test was conducted on all the motion-free scans to assess the benefit of frame conversion on motion correction. To generate motion vector ground truth, we implemented non-rigid motion correction in BioImage Suite \citep{joshi2011unified} (BIS) on an independent $^{82}$Rb cardiac scan set classified as having significant motion by the clinical team. Since the conventional BIS method might fail on early frames with distinct tracer distribution differences, we only utilized the motion field estimates from the late frames with realistic characteristics of simulated motion and then scaled the motion magnitude by 2 as the motion magnitude from those late frames adjacent to the last frame might be too small to show significant motion impact. Before motion simulation, different frame mapping models were applied to all the early frames before frame 27, and then the simulated motion vectors were applied to both original and converted early frames. We applied two motion correction methods: the conventional frame registration using BIS was implemented with the same hyperparameter settings as in \cite{guo2022inter}, and a multiple-frame deep learning registration model (DL) that was well-trained on a whole-body FDG PET dataset as in \cite{guo2022unsupervised}. All the early frames were registered to frame 27 as the reference frame. The frames before and after motion simulation were visualized with overlaid cardiac segmentations as the qualitative evaluation. The mean absolute error is computed as the measurement of motion prediction accuracy, formulated as below,
\begin{equation}
\label{motion_error}
|\Delta \phi| = \frac{1}{T}\sum\limits_{n=1}^{T}(|\phi_{x_n} - \hat{\phi}_{x_n}| + |\phi_{y_n} - \hat{\phi}_{y_n}| + |\phi_{z_n} - \hat{\phi}_{z_n}|)*\frac{1}{3},
\end{equation}
where $T$ is the number of transformation control points in a frame, ($\phi_{x_n}$,$\phi_{y_n}$,$\phi_{z_n}$) is the ground-truth of the motion field, and ($\hat{\phi}_{x_n}$,$\hat{\phi}_{y_n}$,$\hat{\phi}_{z_n}$) is the predicted motion field. The normalized mutual information (NMI) between each moving dynamic frame and the reference frame is also calculated as a measurement of spatial alignment.

In the cross-validation of frame conversion, the motion simulation test was implemented on the 17 test subjects of each fold, resulting in 85 motion simulation cases. 

For the independent dataset with significant motion (Group B), we directly applied frame conversion on the early frames and then the same BIS and DL motion correction methods on both the original and converted frame sequences. Since the motion ground truth is lacking, we only visualized the frames with overlaid segmentations as the qualitative evaluation and calculated frame NMI as the quantitative spatial alignment evaluation. 

For each motion correction group, the motion estimates are directly applied to the motion-corrupted original frames before conversion and without intensity normalization to preserve the tracer activities for the following MBF quantification.

\subsection{Myocardial blood flow analysis}
The three-parameter one-tissue compartment model is fitted using the LVBP TAC as the image-derived input function and the LV myocardium TAC for the estimation of uptake rate constant $K_1$,
\begin{equation}
\label{1tcm}
\frac{dC_t(t)}{dt} = K_1C_p(t) - k_2C_t(t),
\end{equation}
where $C_t(t)$ is the tissue concentration, $C_p(t)$ is the plasma concentration (input function), and $k_2$ is the clearance rate constant. Weighted least squares fitting was applied in the model fitting for parameter estimations, with the weights computed as in \ref{weights} \citep{shi2019direct,shi2021automatic},
\begin{equation}
\label{weights}
w_i = \frac{dur_i^2}{T_i\times DCF^2},
\end{equation}
where for the $i^{th}$ frame, $w_i$ is the fitting weight, $dur_i$ is the frame scan duration, $T_i$ is the sum of the activity of the entire frame, and $DCF$ is the decay correction factor.

The MBF is then quantified from the $K_1$ estimate under the Renkin-Crone Model \citep{renkin1955effects,renkin1959transport,crone1963permeability} with the unit of mL/min/g,
\begin{equation}
\label{mbf}
K_1 = MBF\times (1-\alpha e^{-\beta/MBF}),
\end{equation}
where $\beta$ is the basal permeability-surface area (PS) product and $\alpha$ is the parameter of MBF-related PS changes \citep{germino2016quantification}. Both $\alpha$ and $\beta$ were the same as in the previous $^{82}$Rb study ($\alpha$=0.74, $\beta$=0.51, \citet{shi2021automatic}).

The percentage differences for $K_1$ and $MBF$ are calculated between the motion-corrected and original frames,
\begin{equation}
\label{k1_diff}
{K_1}_{diff} = (\hat{K_1} - K_1)/K_1,
\end{equation}
\begin{equation}
\label{mbf_diff}
MBF_{diff} = (\hat{MBF} - MBF)/MBF,
\end{equation}
where $\hat{K_1}$ and $\hat{MBF}$ are respectively the estimated parameters after motion correction, and $K_1$ and $MBF$ are respectively the estimated parameters before motion correction. For the motion simulation cases, the percentage differences are calculated as estimation bias; for the cases with real motion, since the motion-free ground truth is lacking, the percentage differences are reported as the effect of motion correction.

\section{Results}
\subsection{Frame conversion evaluation}

Figure \ref{conversion_result} displays sample slices from distinct cases of early-to-late frame conversion results under each method. The models of one-to-one conversion were trained with the most precise temporal mapping pairs, but the generation results were generally unsatisfactory with several failure cases. This might be due to the small sample size in the dataset and the possible overfitting of the model since the dataset size of one specific mapping is substantially smaller than the entire dataset. Also, although the model could learn the anatomical information from one early frame, the kinetics-related temporal information from only one early frame is insufficient for satisfactory conversion. The Vanilla GAN was able to learn the conversion patterns in an all-to-one manner, although with some aberrations and distortions in the myocardium. The MSE loss GAN produced results with improved visual similarity. The proposed TAI-GAN achieved the best visual performance with the least amount of remaining mismatch and distortion after incorporating temporal and anatomical information.

Table \ref{frame_quan} summarizes the quantitative results of the image similarity evaluation for frame conversion. Statistics are presented for different sets of test data, including only EQ-1 frames; only EQ+1 frames; all pre-EQ frames, defined as frames where LVBP activity is greater than RVBP activity; and all frames. The quantitative results confirmed that the one-to-one trained models did not perform better than the all-to-one trained models, possibly due to the lack of inter-frame tracer kinetic dependencies. The TAI-GAN produced the best result in each statistic across all evaluation sets. The proposed TAI-GAN showed the greatest improvement in the pre-EQ frames, which are the most challenging subgroup in frame conversion due to the significant tracer distribution discrepancy between the input and output frames.

\begin{figure*}[t]
\centering
\includegraphics[width=0.9\textwidth,keepaspectratio]{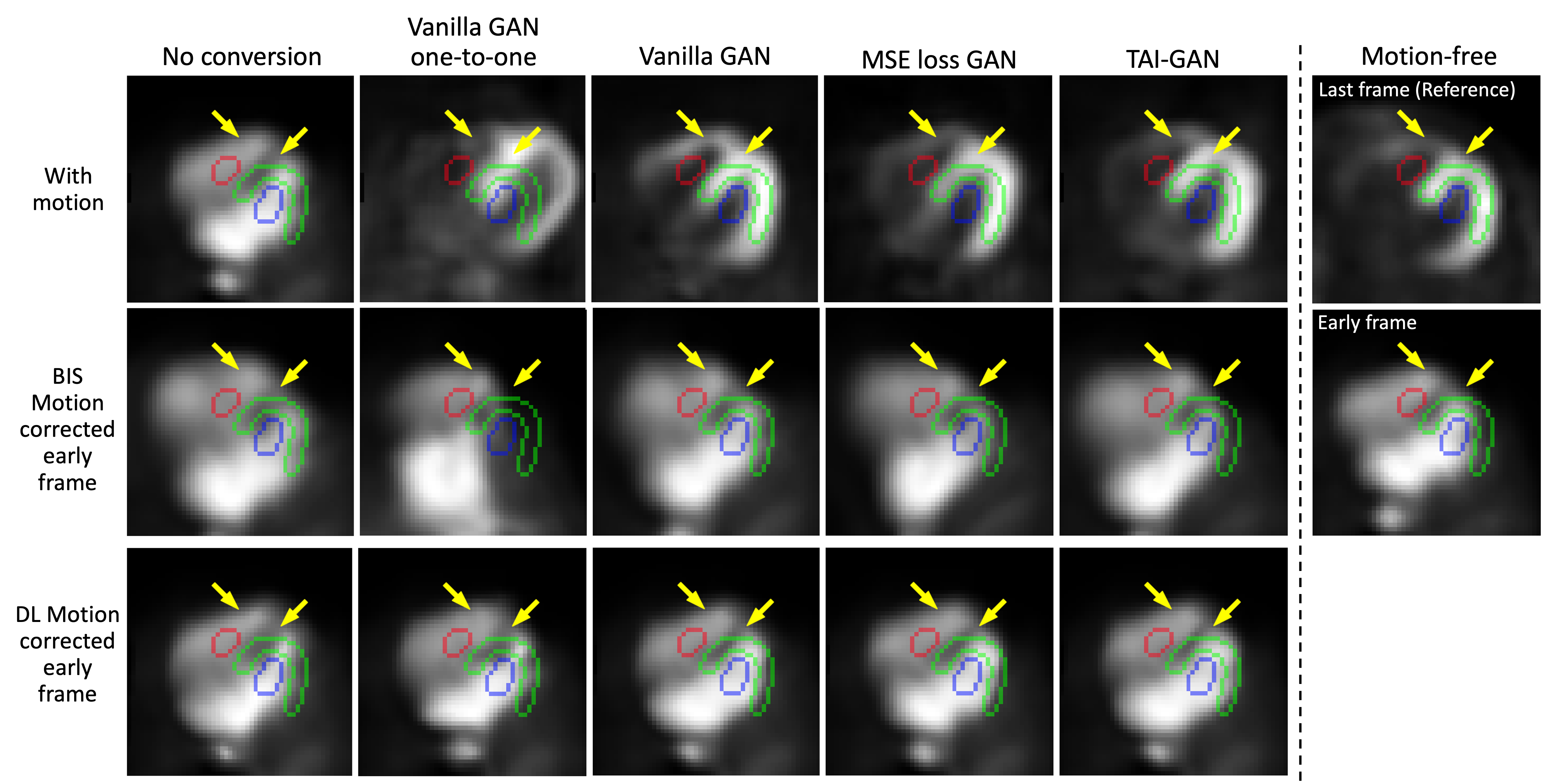}
\caption{Sample motion simulation and correction results with different methods of frame conversion, with overlaid RVBP (red), LVBP (blue), and myocardium (green) segmentation contours and arrows highlighting alignment or mismatch between the cardiac segmentations and structures.}
\label{motion_sim}
\end{figure*}

\subsection{Motion simulation study}
\begin{table*}[h]
\centering
\caption{Mean absolute motion prediction errors without and with each conversion method (in mm, mean $\pm$ standard deviation) with the best results marked \textbf{in bold}.}
\label{tab2}
\resizebox{0.9\textwidth}{!}{
\begin{tabular}{c|c|c|c|c|c|c}
\hline
\tabincell{c}{Motion\\correction\\method} & Test set & \tabincell{c}{No\\Conversion}& \tabincell{c}{Vanilla GAN\\One-to-one} & Vanilla GAN & \tabincell{c}{MSE loss \\GAN } & TAI-GAN\\
\hline
\multirow{3}{*}{BIS} & All frames & 5.30 $\pm$ 0.98$^{*}$ & - & 5.44 $\pm$ 0.90$^{*}$ & 5.06 $\pm$ 0.37$^{*}$ & \textbf{4.70 $\pm$ 0.46}\\
 \cline{2-7}
& EQ-1 & 5.36 $\pm$ 1.52$^{*}$ & 5.42 $\pm$ 1.21 & 5.36 $\pm$ 1.52$^{*}$ & 5.23 $\pm$ 1.10$^{*}$ & \textbf{5.17 $\pm$ 0.60}\\
 \cline{2-7}
& EQ+1 & 5.30 $\pm$ 1.18$^{*}$ & 4.74 $\pm$ 0.93 & 4.93 $\pm$ 1.02$^{*}$ & 4.90 $\pm$ 0.47$^{*}$ & \textbf{4.52 $\pm$ 1.02}\\
\hline
\multirow{3}{*}{DL} & All frames & 3.84 $\pm$ 0.39$^{*}$ & - & 3.82 $\pm$ 0.40$^{*}$ & 3.81 $\pm$ 0.40$^{*}$ & \textbf{3.77 $\pm$ 0.38}\\
 \cline{2-7}
& EQ-1 & 3.87 $\pm$ 0.59$^{*}$ & 3.83 $\pm$ 0.53$^{*}$ & 3.80 $\pm$ 0.60 & 3.80 $\pm$ 0.57 & \textbf{3.78 $\pm$ 0.56}\\
 \cline{2-7}
& EQ+1 & 3.85 $\pm$ 0.53$^{*}$ & 3.79 $\pm$ 0.50$^{*}$ & 3.83 $\pm$ 0.54$^{*}$ & 3.72 $\pm$ 0.56 & \textbf{3.70 $\pm$ 0.54}\\
\hline
\multicolumn{6}{l}{$^{*}P \textless 0.05$ between the current method and TAI-GAN (paired two-tailed t-test).}
\end{tabular}
}
\end{table*}

\begin{table*}[t]
\centering
\caption{Mean frame NMI after motion correction without and with each conversion method (mean $\pm$ standard deviation) with the best results marked \textbf{in bold}.}
\label{frameNMI}
\resizebox{\textwidth}{!}{
\begin{tabular}{c|c|c|c|c|c|c|c|c}
\hline
\tabincell{c}{Motion\\correction\\method} & Test set & No motion & Simulated motion & \tabincell{c}{No\\Conversion}& \tabincell{c}{Vanilla GAN\\One-to-one} & Vanilla GAN & \tabincell{c}{MSE loss \\GAN } & TAI-GAN\\
\hline
\multirow{3}{*}{BIS} & All frames & 0.933 $\pm$ 0.013$^{*}$ & 0.915 $\pm$ 0.015$^{*}$ & 0.887 $\pm$ 0.027$^{*}$ & - & 0.917 $\pm$ 0.035$^{*}$ & 0.918 $\pm$ 0.035$^{*}$ & \textbf{0.948 $\pm$ 0.009}\\
 \cline{2-9}
& EQ-1 & 0.925 $\pm$ 0.015$^{*}$ & 0.914 $\pm$ 0.015$^{*}$ & 0.884 $\pm$ 0.033$^{*}$ & 0.902 $\pm$ 0.033$^{*}$ & 0.906 $\pm$ 0.029$^{*}$ & 0.905 $\pm$ 0.067$^{*}$ & \textbf{0.948 $\pm$ 0.011}\\
 \cline{2-9}
& EQ+1 & 0.927 $\pm$ 0.014$^{*}$ & 0.914 $\pm$ 0.015$^{*}$ & 0.887 $\pm$ 0.034$^{*}$ & 0.902 $\pm$ 0.033$^{*}$ & 0.905 $\pm$ 0.029$^{*}$ & 0.903 $\pm$ 0.081$^{*}$ & \textbf{0.947 $\pm$ 0.010}\\
\hline
\multirow{3}{*}{DL} & All frames & 0.933 $\pm$ 0.013$^{*}$ & 0.915 $\pm$ 0.015$^{*}$ & 0.937 $\pm$ 0.014$^{*}$ & - & \textbf{0.939 $\pm$ 0.013} & 0.938 $\pm$ 0.014$^{*}$ & \textbf{0.939 $\pm$ 0.013}\\
 \cline{2-9}
& EQ-1 & 0.925 $\pm$ 0.015$^{*}$ & 0.914 $\pm$ 0.015$^{*}$ & 0.936 $\pm$ 0.015$^{*}$ & 0.928 $\pm$ 0.018$^{*}$ & 0.938 $\pm$ 0.014$^{*}$ & 0.937 $\pm$ 0.014$^{*}$ & \textbf{0.966 $\pm$ 0.008}\\
 \cline{2-9}
& EQ+1 & 0.927 $\pm$ 0.014$^{*}$ & 0.914 $\pm$ 0.015$^{*}$ & 0.937 $\pm$ 0.015$^{*}$ & 0.929 $\pm$ 0.017$^{*}$ & \textbf{0.939 $\pm$ 0.014} & 0.938 $\pm$ 0.015$^{*}$ & \textbf{0.939 $\pm$ 0.015}\\
\hline
\multicolumn{6}{l}{$^{*}$P \textless 0.05 between the current method and TAI-GAN (paired two-tailed t-test).}
\end{tabular}
}
\end{table*}

The sample results of the motion simulation and the corrected motion are shown in Figure \ref{motion_sim}. In the motion-free sample, the cardiac mask contours are well-aligned with the corresponding anatomical structures. After introducing the simulated non-rigid motion, part of the LVBP is misaligned under the myocardium. When directly registering the original frames without frame conversion, the resliced frame by BIS was more distorted, especially for the LVBP, and the motion correction of DL showed an undercorrected result, possibly because of the substantial tracer distribution differences in the frame pair. In the Vanilla and MSE loss GAN results, the frame generation results look similar to the last reference frame to some extent, but with deformation and shape differences from conversion errors, additional mismatches were observed after BIS motion correction. With the highest frame similarity and minimal frame mapping errors, the conversion result of the proposed TAI-GAN matched the relevant myocardium and ventricle locations with the original early frame and had the cardiac shape most similar to the original last frame. The registration result of BIS demonstrated the best visual alignment of the cardiac ROIs and that of DL showed the best similarity of the base cardiac regions. It is noted that all of the DL motion correction results showed under-correction on top of the LV, possibly due to the gap between the whole-body motion and cardiac motion; still, the result after TAI-GAN conversion reduced this remaining mismatch the most.

Table \ref{tab2} summarizes the mean absolute motion estimation error of both BIS and DL motion correction without conversion and with all the conversion methods. The early scan time of a dynamic frame is generally related to high motion estimation errors, and the DL motion correction method achieved lower estimation errors than the BIS method. In all the included early frames and both EQ-1 and EQ+1 frames, for both BIS and DL motion correction methods, the proposed TAI-GAN achieved the lowest motion estimation error among all the methods. The TAI-GAN-aided BIS motion correction significantly reduced the average prediction error compared to no conversion and all the other GAN models ($p<0.05$), and the TAI-GAN-aided DL motion correction significantly decreased the average prediction error for all frames compared to no conversion and the one-to-one GAN mapping ($p<0.05$). This suggests that the accuracy of motion correction would increase after successful frame conversion.

Table \ref{frameNMI} summarizes the average of frame NMI before and after each motion correction method. For BIS motion correction, the TAI-GAN frame conversion could successfully improve frame similarity measured by NMI compared to no conversion and all the GAN baselines with significant differences ($p<0.05$). For DL motion corrections, even without conversion, the frame NMI increased. For the EQ-1 frame, the TAI-GAN significantly improved frame NMI compared with all the other baselines ($p<0.05$), but for the EQ+1 frame and all the frames the TAI-GAN achieved comparable improvement as Vanilla GAN ($p>0.05$). It could be due to the limitation of the direct application of a well-trained whole-body FDG model to the cardiac Rb-82 data, especially for later frames with less tracer distribution difference, where the motion characteristics-related generalization gap became more significant.

\subsection{Parametric fitting results}
\begin{table}[t]
\centering
\caption{$K_1$ and MBF quantification results after BIS or DL motion correction (mean $\pm$ standard deviation) with the best results marked \textbf{in bold}.}
\label{tab3}
\resizebox{0.48\textwidth}{!}{
\begin{tabular}{c|c|c|c}
\hline
& \tabincell{c}{Mean $K_1$ percentage\\difference (\%)}& \tabincell{c}{Mean MBF percentage\\difference (\%)} & \tabincell{c}{Mean fitting\\error ($\times$10$^{-5}$) } \\
\hline
Motion-free & - & - & 1.80 $\pm$ 2.06 \\
\hline
With motion & 15.72 $\pm$ 26.24$^{*}$ & 4.43 $\pm$ 38.67$^{*}$ & 10.46 $\pm$ 6.55$^{*}$\\
\hline
BIS & 16.63 $\pm$ 32.13$^{*}$ & 14.45 $\pm$ 68.65$^{*}$ & 13.74 $\pm$ 19.49$^{*}$\\
\hline
Vanilla GAN+BIS & 11.73 $\pm$ 32.82$^{*}$ & 3.09 $\pm$ 39.09 & 9.60 $\pm$ 9.02$^{*}$\\
\hline
MSE loss GAN+BIS & 17.98 $\pm$ 24.54$^{*}$ & 3.32 $\pm$ 31.76 & 11.34 $\pm$ 12.29$^{*}$\\
\hline
TAI-GAN+BIS & \textbf{6.42 $\pm$ 35.82} & \textbf{2.21 $\pm$ 35.41} & \textbf{7.41 $\pm$ 7.81}\\
\hline
\hline
DL & 17.43 $\pm$ 23.89$^{*}$ & 7.03 $\pm$ 40.37$^{*}$ & 8.35 $\pm$ 7.52$^{*}$\\
\hline
Vanilla GAN+DL & 17.34 $\pm$ 24.38$^{*}$ & 7.87 $\pm$ 40.82$^{*}$ & 8.29 $\pm$ 7.58$^{*}$\\
\hline
MSE loss GAN+DL & 18.54 $\pm$ 24.12$^{*}$ & 5.90 $\pm$ 32.36 & 7.97 $\pm$ 7.26\\
\hline
TAI-GAN+DL & \textbf{15.43 $\pm$ 26.04} & \textbf{3.40 $\pm$ 32.90} & \textbf{7.80 $\pm$ 6.77}\\
\hline
\multicolumn{4}{l}{$^{*}$P $\textless 0.05$ between the current method and TAI-GAN (paired two-tailed t-test).}\\
\end{tabular}
}
\end{table}

\begin{figure*}[t]
\centering
\includegraphics[width=0.85\textwidth,keepaspectratio]{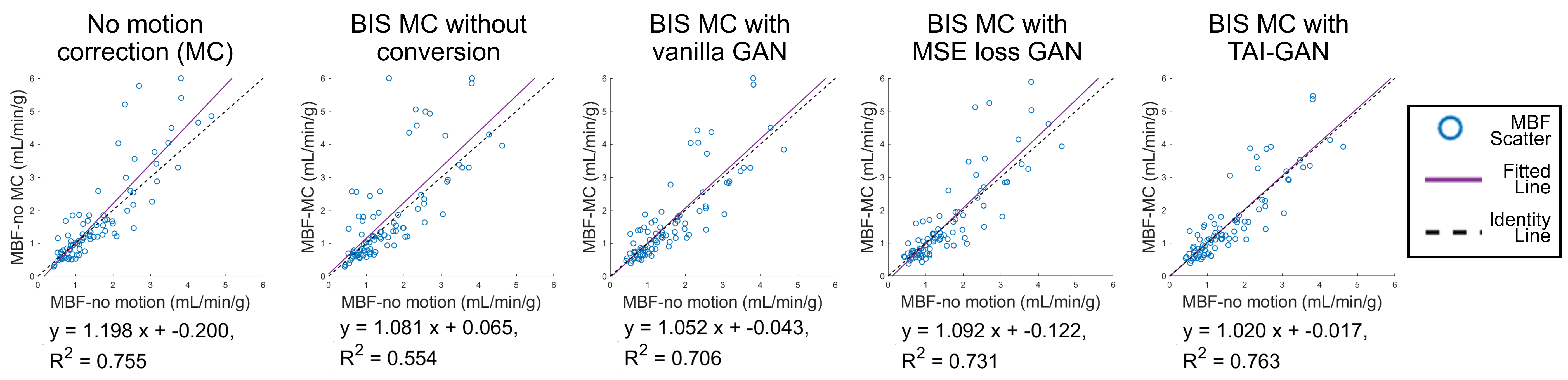}
\caption{Scatter plots of MBF results estimated from no motion frames vs.\ no motion correction and BIS motion correction after different conversion methods.}
\label{MBF}
\end{figure*}

\begin{figure*}[t]
\centering
\includegraphics[width=1\textwidth,keepaspectratio]{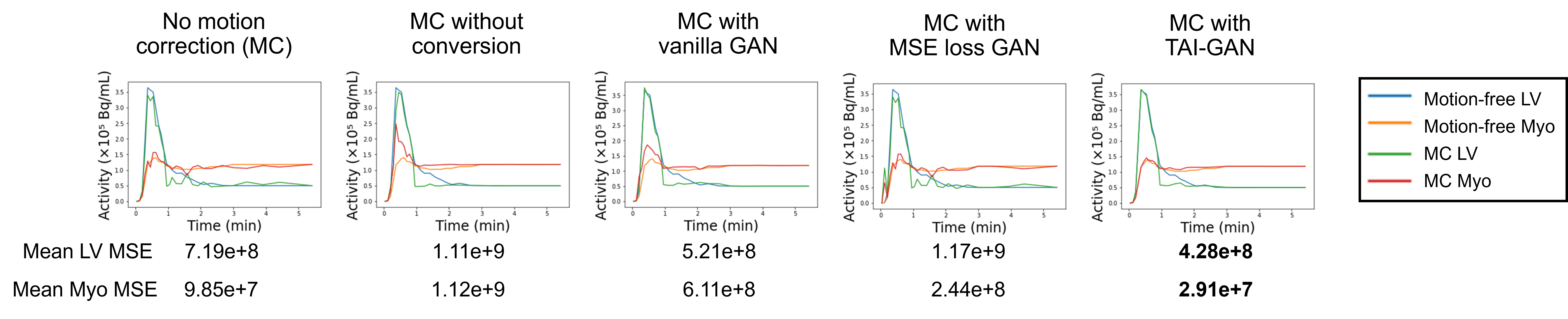}
\caption{A comparison of LVBP and myocardium TACs of each conversion method before and after BIS motion correction.}
\label{TACs}
\end{figure*}

Table \ref{tab3} summarizes the parametric fitting error as well as the percentage difference in $K_1$ and MBF compared to no motion ground truth. The fitting errors of both TAI-GAN+BIS and TAI-GAN+DL were the lowest among all the test classes in each motion correction category, showing significant improvements compared to fitting using frames with motion and all the motion correction methods ($p<0.05$). The $K_1$ and MBF percentage differences of TAI-GAN+BIS and TAI-GAN+DL were the lowest among all the groups. The TAI-GAN+BIS achieved $K_1$ and MBF fitting results closer to those of the motion-free baseline than the TAI-GAN+DL. Thus, we visualized the scatter plots of MBF fitting and the TAC plots of the TAI-GAN+BIS. As in Figure \ref{MBF}, with simulated motion, the scatters of MBF estimates versus the ground truth without motion were not uniformly distributed and were distant from the identity line. With BIS MC but without conversion, the fitted line was close to the identity line but for some data points, the estimation bias even increased. The fitted lines of motion correction with vanilla GAN and MSE loss GAN were closer to the identity line but still with an under-estimation effect. The fitted line of motion correction with the proposed TAI-GAN was the closest to the identity line, suggesting the most improvement in MBF quantification. As in Figure \ref{TACs}, after BIS MC with no conversion, the discrepancy between the two LVBP TACs decreased in later frames but drastically increased in the very early blood pool phase frames. The Vanilla GAN conversion improved the correction of later frames but the early frames remained mismatched in the myocardium. The MSE loss GAN corrected the later frames but showed remaining errors of under-correction or miscorrection in the early and later frames. The TACs of TAI-GAN-aided BIS motion correction were closest to the no-motion ground truth and resulted in the lowest LV and myocardium MSE. 

\subsection{Real-motion evaluation}
\begin{figure*}[t]
\centering
\includegraphics[width=0.85\textwidth,keepaspectratio]{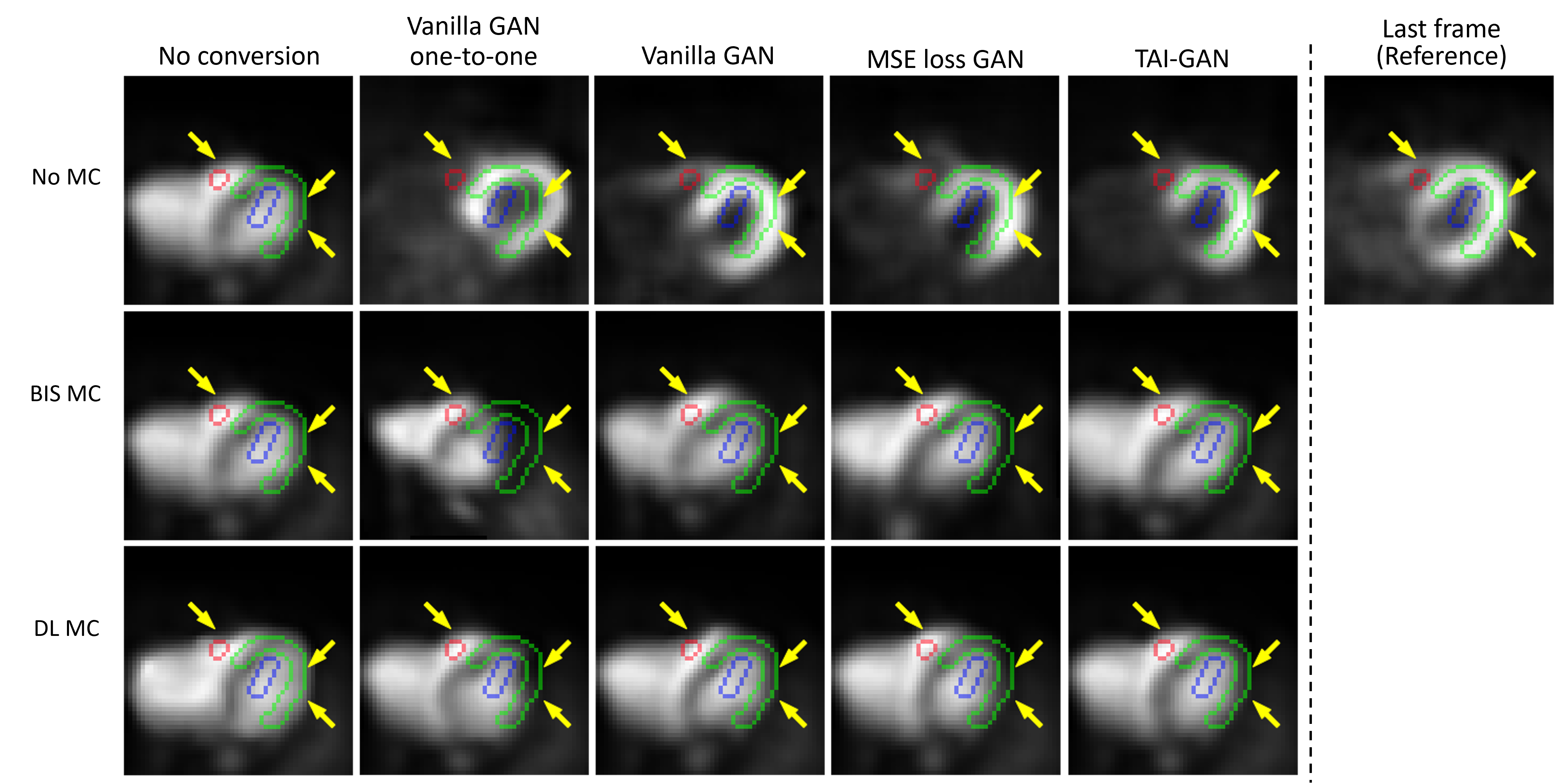}
\caption{Sample patient real motion correction results with different methods of frame conversion, with overlaid RVBP (red), LVBP (blue), and myocardium (green) segmentation contours and arrows highlighting alignment or mismatch between the cardiac segmentations and structures.}
\label{motion_real}
\end{figure*}

A sample of the real-motion correction after frame conversion is shown in Figure \ref{motion_real}. The early and late frames are misaligned in the right ventricle, left ventricle, and myocardium, with non-rigid motion. Without frame conversion, the BIS method showed under correction and the DL method showed motion estimation error in the myocardium region. The one-to-one model conversion showed large errors and distortion, resulting in over-correction after applying both BIS and DL. Both all-to-one baseline models produced more realistic results but still with local spatial distortion. In the conversion result of TAI-GAN, the myocardium shape is the closest to the reference with the location matching with the motion-present early frame. The motion correction result after TAI-GAN conversion was also the best among both the BIS and DL methods.

\begin{table*}[t]
\centering
\caption{Mean frame NMI after motion correction without and with each conversion method (mean $\pm$ standard deviation) on the real-patient motion dataset with the best results marked \textbf{in bold}.}
\label{frameNMI_realmotion}
\resizebox{\textwidth}{!}{
\begin{tabular}{c|c|c|c|c|c|c|c}
\hline
\tabincell{c}{Motion\\correction\\method} & Test set & No motion correction & \tabincell{c}{No\\Conversion}& \tabincell{c}{Vanilla GAN\\One-to-one} & Vanilla GAN & \tabincell{c}{MSE loss \\GAN } & TAI-GAN\\
\hline
\multirow{3}{*}{BIS} & All frames & 0.970 $\pm$ 0.012$^{*}$ & 0.945 $\pm$ 0.013$^{*}$ & - & 0.939 $\pm$ 0.032$^{*}$ & 0.935 $\pm$ 0.037$^{*}$ & \textbf{0.961 $\pm$ 0.008}\\
 \cline{2-8}
& EQ-1 & 0.960 $\pm$ 0.014$^{*}$ & 0.940 $\pm$ 0.012$^{*}$ & 0.864 $\pm$ 0.031$^{*}$ & 0.906 $\pm$ 0.045$^{*}$ & 0.905 $\pm$ 0.052$^{*}$ & \textbf{0.961 $\pm$ 0.009}\\
 \cline{2-8}
& EQ+1 & \textbf{0.964 $\pm$ 0.011} & 0.942 $\pm$ 0.012$^{*}$ & 0.856 $\pm$ 0.024$^{*}$ & 0.924 $\pm$ 0.028$^{*}$ & 0.920 $\pm$ 0.038$^{*}$ & 0.962 $\pm$ 0.011\\
\hline
\multirow{3}{*}{DL} & All frames & 0.970 $\pm$ 0.012$^{*}$ & 0.982 $\pm$ 0.006$^{*}$ & - & 0.982 $\pm$ 0.006$^{*}$ & 0.982 $\pm$ 0.006$^{*}$ & \textbf{0.988 $\pm$ 0.004}\\
 \cline{2-8}
& EQ-1 & 0.960 $\pm$ 0.014$^{*}$ & 0.981 $\pm$ 0.006$^{*}$ & 0.981 $\pm$ 0.006$^{*}$ & 0.981 $\pm$ 0.006$^{*}$ & 0.981 $\pm$ 0.006$^{*}$ & \textbf{0.988 $\pm$ 0.004}\\
 \cline{2-8}
& EQ+1 & 0.964 $\pm$ 0.011$^{*}$ & 0.982 $\pm$ 0.006$^{*}$ & 0.982 $\pm$ 0.006$^{*}$ & 0.982 $\pm$ 0.006$^{*}$ & 0.982 $\pm$ 0.006$^{*}$ & \textbf{0.988 $\pm$ 0.004}\\
\hline
\multicolumn{6}{l}{$^{*}$P \textless 0.05 between the current method and TAI-GAN (paired two-tailed t-test).}
\end{tabular}
}
\end{table*}

Quantitative evaluations of real-motion cases are summarized in Table \ref{frameNMI_realmotion}. Since motion ground truth is not available in the real-motion cases, frame similarity as measured by NMI was calculated to assess motion correction results. Similar to the motion simulation test, the DL motion correction achieved higher frame NMI than the BIS method. For the BIS method, the TAI-GAN conversion improved motion correction the best in all frames and the EQ-1 frame set. For the EQ+1 set, though all the methods produced lower frame NMI than the original frames, the TAI-GAN achieved the average frame NMI closest to the original frames and the difference was not significant ($p<0.05$). For the DL method, the motion correction results without conversion did not show a significant difference compared with other frame conversion baselines. TAI-GAN conversion results showed significant improvements compared to all the baselines.

\begin{table}[t]
\centering
\caption{$K_1$ and MBF quantification results after BIS or DL motion correction (mean $\pm$ standard deviation) on the real-motion dataset with the best results marked \textbf{in bold}.}
\label{realmotion_mbf}
\resizebox{0.49\textwidth}{!}{
\begin{tabular}{c|c|c|c}
\hline
& \tabincell{c}{Mean $K_1$ percentage\\difference (\%)}& \tabincell{c}{Mean MBF percentage\\difference (\%)} & \tabincell{c}{Mean fitting\\error ($\times 10^{-4}$) } \\

\hline
Without MC & - & - & 2.53 $\pm$ 2.72$^{*}$\\
\hline
BIS & -5.23 $\pm$ 15.52 & -0.56 $\pm$ 7.75 & 3.48 $\pm$ 7.19$^{*}$\\
\hline
Vanilla GAN+BIS & -17.61 $\pm$ 11.41 & -6.99 $\pm$ 17.30 & 1.96 $\pm$ 2.70$^{*}$\\
\hline
MSE loss GAN+BIS & -20.98 $\pm$ 13.48 & -6.85 $\pm$ 21.87 & 2.53 $\pm$ 3.26$^{*}$\\
\hline
TAI-GAN+BIS & -12.20 $\pm$ 12.63 & -6.96 $\pm$ 15.61 & \textbf{1.21 $\pm$ 1.98}\\
\hline
\hline
DL & -21.77 $\pm$ 35.52 & -25.40 $\pm$ 43.95 & 4.07 $\pm$ 5.07$^{*}$\\
\hline
Vanilla GAN+DL & -27.42 $\pm$ 32.31 & -33.65 $\pm$ 39.85 & 2.18 $\pm$ 2.55$^{*}$\\
\hline
MSE loss GAN+DL & -33.06 $\pm$ 30.47 & -40.91 $\pm$ 34.30 & 1.74 $\pm$ 2.02$^{*}$\\
\hline
TAI-GAN+DL & -27.46 $\pm$ 32.36 & -33.16 $\pm$ 37.76 & \textbf{1.54 $\pm$ 2.17}\\
\hline
\multicolumn{4}{l}{$^{*}$P $\textless 0.05$ between the current method and TAI-GAN (paired two-tailed t-test).}\\
\end{tabular}
}
\end{table}

Compared to the no motion correction real-motion dataset results, the parametric fitting error as well as the percentage difference in both $K_1$ and MBF are summarized in Table \ref{realmotion_mbf}. The DL showed more substantial differences in $K_1$ and MBF estimations than BIS. Compared to no conversion motion correction results, frame conversion also showed a more substantial difference in both motion correction methods. The TAI-GAN conversion resulted in significantly reduced fitting errors after either BIS or DL motion correction.

\begin{figure*}[t]
\centering
\includegraphics[width=0.85\textwidth,keepaspectratio]{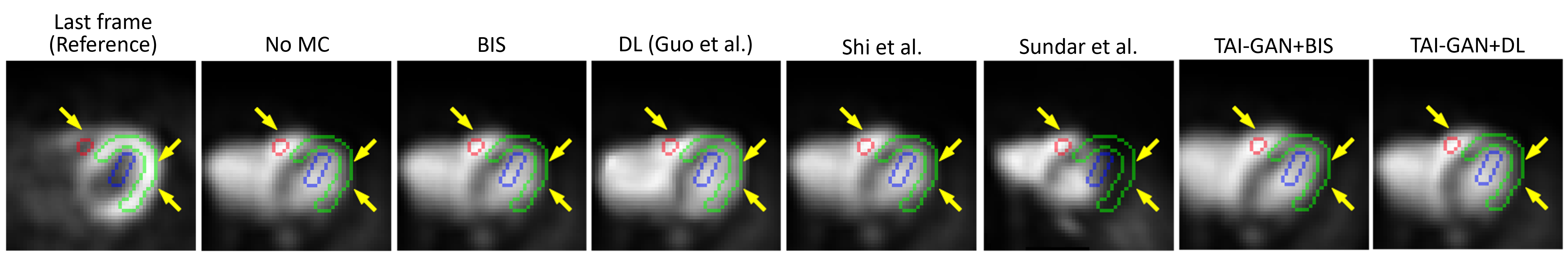}
\caption{A comparison of the current motion correction baselines on real-motion cases, with overlaid RVBP (red), LVBP (blue), and myocardium (green) segmentation contours and arrows highlighting alignment or mismatch between the cardiac segmentations and structures.}
\label{mc_baseline}
\end{figure*}

A visual comparison of all the current motion correction pipelines on the real-motion dataset is shown in Figure \ref{mc_baseline}. The BIS method failed to correct the misalignment. When well-trained on the whole-body FDG dataset and then directly applied to the cardiac dataset, the DL method showed its limitation in introducing new distortion to the myocardium region. The method of \citet{shi2021automatic} was able to correct the location-wise misalignment, but the method was only for translational motion correction, and the shape of the left ventricle was not perfectly aligned with the myocardium mask outline. Also, note that Shi et al.\ trained two separate models for the early and late frames, resulting in additional time and memory consumption. Among all the frame-conversion-aided methods, the method of \citet{sundar2021conditional} failed, producing the most unrealistic motion prediction, possibly due to the errors in the one-to-one frame conversion method. The TAI-GAN+BIS corrected the motion in the right ventricle better, while the TAI-GAN+DL corrected the mismatch in the left ventricle and myocardium better. With TAI-GAN frame conversion, both the visual performance of BIS and DL motion correction improved. 

\subsection{Ablation studies of temporal and anatomical information}

\begin{figure*}[t]
\centering
\includegraphics[width=0.95\textwidth,keepaspectratio]{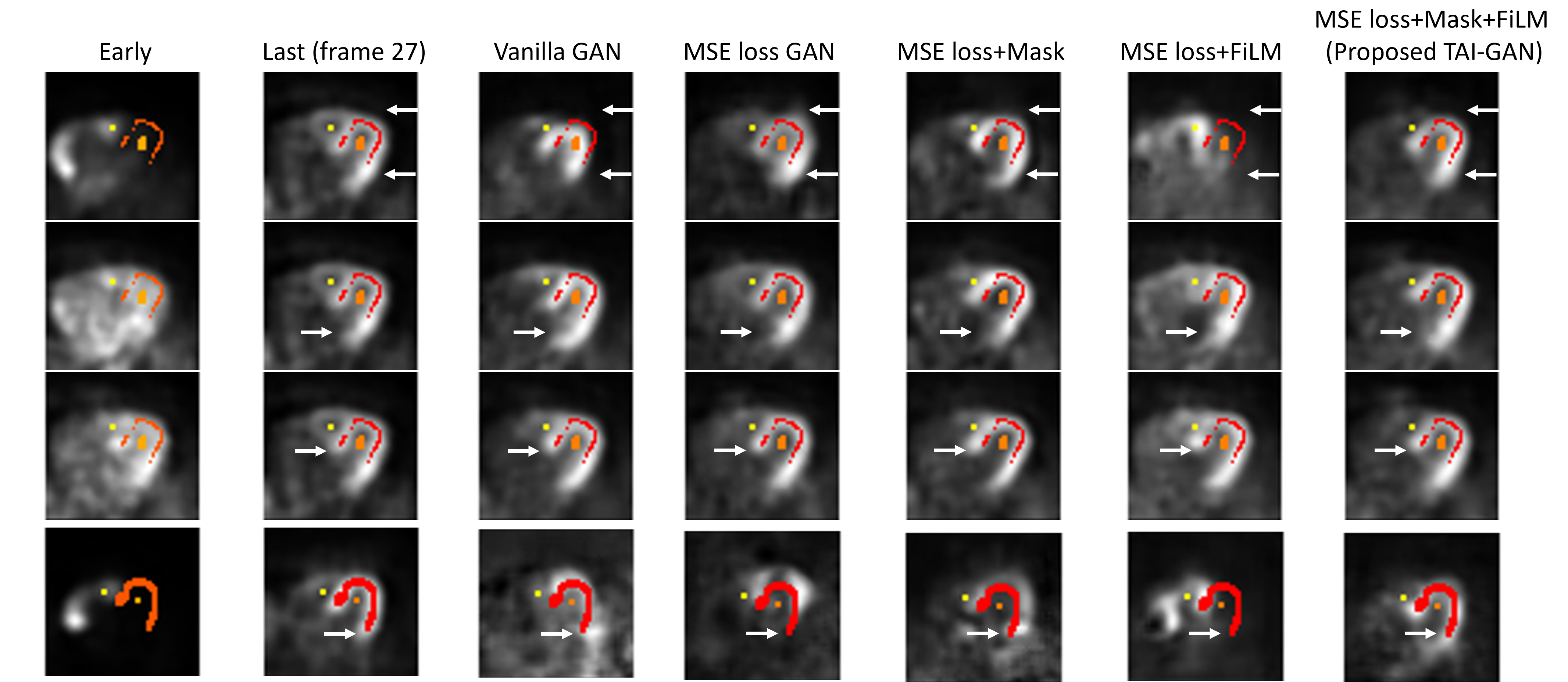}
\caption{Sample results of the ablation study as the evaluation of the introduced temporal and anatomic information with overlaid RVBP (yellow), LVBP (orange), and myocardium (red) segmentations.}
\label{ablation}
\end{figure*}

Figure \ref{ablation} shows sample results with overlaid cardiac segmentations from the ablation study as the evaluation of the introduced temporal and anatomic information. The auxiliary anatomical information provided by the cardiac masks further improved frame conversion in addition to the MSE loss, but some of the misrepresentations related to insufficient temporal interdependence still exist in the conversion results. The introduced temporal input informed the model effectively with the kinetics information, especially improving model performance in post-EQ frames, but it is noted that due to the lack of anatomical information, for pre-EQ frame samples, the model might misidentify the RVBP as LVBP and then cause incorrect myocardium prediction. 

Table \ref{ablation} summarizes the quantitative assessments of the complete ablation studies. The anatomical information provided by the rough segmentation masks showed substantial improvement in both all frames and all pre-EQ frame test sets. The temporal information introduced by TACs and the FiLM model further improved SSIM and PSNR on all the frames compared to MSE loss GAN, although on all the pre-EQ frames the performance dropped possibly due to the lack of anatomical guidance. With the combined auxiliary anatomical and temporal information, the proposed TAI-GAN achieved the best performance.

\begin{table*}[t]
\centering
\caption{Quantitative image similarity evaluation results (mean $\pm$ standard deviation) of the ablation study of the introduced temporal and anatomic information with the best results marked \textbf{in bold}.}
\label{ablation}
\resizebox{0.8\textwidth}{!}{
\begin{tabular}{c|c|c|c|c|c|c}
\hline
Test set & Metric & Vanilla GAN & MSE loss & MSE loss+Mask & MSE loss+FiLM & \tabincell{c}{MSE loss+Mask+FiLM \\(Proposed TAI-GAN)}\\
\hline

\multirow{4}{*}{\tabincell{c}{All\\Pre-EQ\\frames}}  & NMAE & 0.072 $\pm$ 0.007$^{*}$ & 0.066 $\pm$ 0.011 & 0.065 $\pm$ 0.008$^{*}$ & 0.079 $\pm$ 0.009$^{*}$ & \textbf{0.062 $\pm$ 0.008}\\
 \cline{2-7}
& MSE & 0.063 $\pm$ 0.010$^{*}$ & 0.053 $\pm$ 0.016 & 0.049 $\pm$ 0.010$^{*}$ & 0.081 $\pm$ 0.014$^{*}$ & \textbf{0.046 $\pm$ 0.009}\\
 \cline{2-7}
 & SSIM & 0.594 $\pm$ 0.012$^{*}$ & 0.596 $\pm$ 0.047$^{*}$ & 0.621 $\pm$ 0.033 & 0.587 $\pm$ 0.021$^{*}$ & \textbf{0.627 $\pm$ 0.025}\\
 \cline{2-7}
 & PSNR & 18.51 $\pm$ 0.47$^{*}$ & 19.27 $\pm$ 1.04$^{*}$ & 19.63 $\pm$ 0.72 & 17.42 $\pm$ 0.85$^{*}$ & \textbf{19.83 $\pm$ 0.74}\\
 \hline
  \hline
  
 \multirow{4}{*}{\tabincell{c}{All \\frames}} & NMAE & 0.047 $\pm$ 0.004$^{*}$ & 0.044 $\pm$ 0.002 & 0.042 $\pm$ 0.002 & 0.045 $\pm$ 0.003$^{*}$ & \textbf{0.040 $\pm$ 0.002}\\
 \cline{2-7}
 & MSE & 0.027 $\pm$ 0.002$^{*}$ & 0.024 $\pm$ 0.003 & \textbf{0.021 $\pm$ 0.002} & 0.026 $\pm$ 0.003$^{*}$ & \textbf{0.021 $\pm$ 0.002}\\
 \cline{2-7}
 & SSIM & 0.708 $\pm$ 0.010$^{*}$ & 0.716 $\pm$ 0.007$^{*}$ & 0.720 $\pm$ 0.015$^{*}$ & 0.723 $\pm$ 0.018$^{*}$ & \textbf{0.733 $\pm$ 0.018}\\
 \cline{2-7}
 & PSNR & 22.80 $\pm$ 0.53$^{*}$ & 23.24 $\pm$ 0.34$^{*}$ & 23.57 $\pm$ 0.28 & 23.40 $\pm$ 0.36 & \textbf{23.80 $\pm$ 0.47}\\
 \hline
\multicolumn{6}{l}{{\footnotesize  $^{*}$P \textless 0.05 between the current method and TAI-GAN (subject-wise paired two-tailed t-test).}}\\ 

\end{tabular}
}
\end{table*}

\section{Discussion}
In this study, we propose the Temporally and Anatomically Informed GAN (TAI-GAN), a new framework that uses an all-to-one mapping method to translate the appearance of all early frames into the last reference frame to tackle the tracer discrepancy problem in cardiac dynamic PET motion correction. We use a FiLM layer to encode tracer kinetics-related temporal information and rough cardiac structural segmentations to provide anatomical guidance. We evaluated TAI-GAN in terms of frame conversion similarity, motion correction accuracy, and MBF quantification errors using the five-fold cross-validation. On the clinical $^{82}$Rb PET dataset, the TAI-GAN achieved desirable frame conversion and improved current motion correction methods, as well as the downstream clinical MBF quantification.

Currently, the anatomical information is provided by the cardiac ROIs as an additional input channel. Note that in the current workflow, the need for cardiac ROIs did not introduce additional labor since these cardiac segmentations are already required in MBF quantification and constitute an essential part of the current clinical workflow. In future work, we will further automate this step by applying existing automatic labeling using the clinical software 4DM (INVIA, Ann Arbor, Michigan; \citet{ficaro2007corridor4dm}) or deep learning segmentation models. Given the challenge of acquiring manual dense annotations as the labels for fully supervised training, weakly-supervised models have been widely investigated in the field of medical image segmentation \citep{wang2019weakly,guo2023seam}. Also, utilizing non-contrast CT images as additional guidance is another future direction to achieve a segmentation-free method.

Another potential limitation of the current work is that the frame conversion method is developed using a clinical dataset that is identified by clinicians as nearly motion-free. Since the cardiac ROIs are segmented based on the last reference frame, the mismatch between the last frame ROI and the early frames might introduce errors in frame conversion if motion is present. To overcome this problem, we implemented random translation of cardiac ROIs during training. Thus, the network is trained to handle any present mismatch between the cardiac ROIs and the dynamic frames and only takes the anatomical input as auxiliary guidance instead of fully relying on that information. This is particularly useful in distinguishing between the early RV and LV phases and preventing the model from predicting myocardium structure near the RV instead of LV. Also, the residual minimal motion in the nearly motion-free dataset might introduce errors in the motion simulation tests. Further investigations including simulation studies on phantoms with motion ground truth are warranted.

A comparison of the memory footprint and the average training time for each frame conversion method is summarized in Supplementary Table \ref{memory}. The average training time for all the mappings needed to convert a whole sequence using the vanilla GAN one-to-one model is the longest among all the frame conversion methods and is ~10 times longer than the all-to-one mapping methods, making it infeasible to implement in practice. The training time of the TAI-GAN is still comparable to the other two baselines, indicating the efficiency of the incorporation of temporal and anatomical information. With an additional channel, the memory footprint of the TAI-GAN is ~1.5 times of other baselines.

There are several limitations and future directions of this study. First, this method is implemented as a post-reconstruction method, taking only the inter-frame motion into consideration. The attenuation mismatch and intra-frame motion also degrade the quality of the reconstructed frames and introduce errors to the MBF quantification. Future work includes correcting intra-frame and attenuation mismatches in the workflow. Second, the current work only includes ROI-based MBF quantification as clinical analysis. A future direction for clinical-centric evaluation is validating the downstream clinical impact using invasive catheterization as the clinical gold standard. Including post hoc analysis of diagnostic accuracy and clinical experts' reading reports is also worth further investigation. Third, current predictions of the late frames serve as the intermediate results as the input to the following motion correction step. These outputs have comparable visual characteristics to the real late frames but do not represent the actual tracer activities. This is due to the intensity normalization in model training to benefit the optimization process. Future work includes directly predicting motion-compensated late frames with the actual intensities. This will reduce scan time and further simplify the current protocol. Besides, currently, temporal information is sent to the generator indirectly through encoding and FiLM. Future work will investigate introducing more direct kinetics-based optimization \citep{guo2022mcp,guo2023mcp} to the network to provide more realistic predictions. Finally, we will investigate other advanced generative backbone structures such as diffusion models \citep{ho2020denoising} and neural ordinary differential equations \citep{chen2018neural} to further improve conversion performance.

\section{Conclusion}
In conclusion, we propose a novel Temporally and Anatomically Informed GAN (TAI-GAN) as an all-to-one mapping method to convert all the early frames to the last frame to address the challenge of fast tracer distribution change in cardiac dynamic PET inter-frame motion correction. The tracer kinetics-related temporal information is incorporated by a FiLM layer into the generator bottleneck, and cardiac segmentations are concatenated with the early frame for network input to provide anatomical guidance. Under a 5-fold cross-validation on the clinical $^{82}$Rb cardiac dataset, the TAI-GAN achieved desirable frame conversion results with comparable visual quality to the real late frames and the highest quantitative frame similarity measurements. In both a motion simulation test and a real motion evaluation, the proposed TAI-GAN increased the prediction accuracy of current motion correction methods and demonstrated improvement in the MBF quantification.

\section*{Declaration of competing interests}
Bruce Spottiswoode is an employee of Siemens Medical Solutions USA, Inc. There are no additional financial conflicts of interest that may be relevant to this paper.

\section*{CRediT authorship contribution statement}
\textbf{Xueqi Guo}: Conceptualization, Methodology, Software, Visualization, Validation, Formal analysis, Investigation, Data Curation, Writing - original draft, review \& editing.
\textbf{Luyao Shi}: Data Curation, Visualization, Writing - review \& editing.
\textbf{Xiongchao Chen}: Data Curation, Validation, Writing - review \& editing.
\textbf{Qiong Liu}: Methodology, Writing - review \& editing.
\textbf{Bo Zhou}: Methodology, Writing - review \& editing.
\textbf{Huidong Xie}: Methodology, Writing - review \& editing.
\textbf{Yi-Hwa Liu}: Data Curation, Writing - review \& editing.
\textbf{Richard Palyo}: Data Curation, Writing - review \& editing.
\textbf{Edward J. Miller}: Writing - review \& editing.
\textbf{Albert J. Sinusas}: Writing - review \& editing.
\textbf{Lawrence Staib}: Writing - review \& editing.
\textbf{Bruce Spottiswoode}: Conceptualization, Writing - review \& editing, Funding acquisition.
\textbf{Chi Liu}: Conceptualization, Methodology, Writing - review \& editing, Supervision, Funding acquisition.
\textbf{Nicha C. Dvornek}: Conceptualization, Methodology, Writing - review \& editing, Supervision.

\section*{Acknowledgments}
This work was supported by the National Institutes of Health (NIH) under grant R01 CA224140. 

An early version of this work was initially accepted for publication by Springer Nature at SASHIMI: Simulation and Synthesis in Medical Imaging, a workshop of the International Conference on Medical Image Computing and Computer-Assisted Intervention (MICCAI) 2023 \citep{guo2023sashimi}. Here, we included a more comprehensive analysis of the complete evaluation using simulated motion and ablation studies, a comparison of the conventional non-rigid and deep learning-based motion correction methods, an evaluation of an independent cohort of real-patient motion cases, additional background and implementation details, and in-depth discussions considering the viability of clinical use and prospective future directions.

\bibliographystyle{model2-names.bst}\biboptions{authoryear}
\bibliography{refs}

\section*{Supplementary Material}
\setcounter{table}{0} 
\renewcommand\thefigure{S\arabic{figure}} 
\renewcommand\thetable{S\arabic{table}} 
\begin{table}[!h]
\centering
\caption{A comparison of the memory footprint and the average training time for each frame conversion method.}
\label{memory}
\resizebox{0.48\textwidth}{!}{
\begin{tabular}{c|c|c}
\hline
Method & \tabincell{c}{Memory \\footprint (MB)} & \tabincell{c}{Average training time for\\ all frame conversions (min)} \\
\hline
\tabincell{c}{Vanilla GAN \\one-to-one} & 2671 & 780 \\
Vanilla GAN & 2671 & 89 \\
MSE loss GAN & 2671 & 78 \\
TAI-GAN (Proposed) & 3759 & 98 \\
\hline
\end{tabular}
}
\end{table}

\end{document}